# New approach for obtaining ceramic NASICON (Na$_3$Zr$_2$(SiO$_4$)$_2$PO$_4$) films sintered in situ by a sol-gel method, using spray deposition and Near-Infra Red Sintering

Rafael Marti Valls,[a*] Rebecca Griffin,[a*] Anne Sawhney,[a] Celina Domingos-Dlofo,[a] Tom Dunlop,[a] Sam Reis[a], Peter J. Holliman[a], Jenny Baker,[a]

[a]Swansea University Bay Campus, Fabian Way, Swansea, UK

*Contributed the same to the communication authorship

Abstract

In this work we demonstrate a NASICON film sintered in situ onto a fused silica substrate. This production method drastically reduces the manufacturing time by combining the use of a spray-coated sol-gel solution and near-infrared (NIR) ultrafast sintering technology. The first demonstration of NIR sintered ceramics at high temperatures (~1000 °C).

Communication

Sodium super ionic conductors (NASICON) have been widely studied thanks to their unique properties for applications such as harmful gas sensing (e.g. NO$_x$, CO$_2$, SO$_2$ or H$_2$S) [1,2], ionic selective membranes [3,4] or as solid electrolytes in sodium-ion batteries (SIB) [5–7]. Depending on the application, modifications on the structure have been tested to increase conductivity or reduce detrimental secondary phases, among others [7–9]. Several synthetic routes have been studied to produce NASICON structures [10] where solid-state [5,6,11,12], co-precipitation [13–15] and sol-gel [1,13,16–20] methods stand out. Solid state methods present a lower chemical preparation complexity than the co-precipitation and sol-gel methods. However, the processing, multiple ball milling and pressing steps where yield loss at each step can be significant, combined with multiple sintering steps significantly increase production time for solid-state (SS) produced NASICON discs/pellets. Thin (86 micron thick) stand-alone films (precursor powder prepared through SS method) have been demonstrated in batteries with a capacity of 73mAhg$^{-1}$ [21], but replicating thin stand-alone ceramic electrolytes on large area batteries will prove difficult due to the fragility of the stand-alone film.

By comparison, the sol-gel method requires a more sophisticated processing route to obtain the appropriate chemical precursors, the application of different temperatures during the production process, a specific order of addition for the chemicals or the control of their addition rate [22]. This extra care in the manufacture of precursors yields more homogeneous materials and leads shorter sintering times (~ 15 hours compared with >40 hours), which results in the method being more sustainable in terms of energy consumption [23]. The sol preparation process can also be simplified by using water instead of an organic solvent, e.g., ethanol [22]. Shimizu et al. designed a water-based sol-gel method to obtain high-purity Na$_3$Zr$_2$Si$_2$PO$_{12}$ powders by mixing the inorganic precursors dissolved in a certain order [17]. They also developed a method to prepare NASICON films directly from the precursors mixture by modifying the previously mentioned aqueous solution with an hydroxyacid to increase the solution viscosity and stability and subsequently depositing it by spin-coating [18]. The films required 15 hours (5x3hours) of sintering at 1000 °C and produced NASICON films on alumina with some zirconia impurities. This method allows obtaining a sol that is sufficiently stable. A modification of this method was used in this work. Specifically, we fabricated NASICON films by spray deposition instead of spin coating. Spray deposition of other ceramic materials has been shown to substantially simplify the production process, facilitate scalability and enhance controllability [24].

Another limiting step in the NASICON production is the long times of sintering at high temperature typically using radiant heat ovens which can cause secondary phases such as zirconia [18]. An alternative to greatly reduce sintering times is heating with near infrared (NIR) radiation. This technology was studied in the present work. Whilst NIR has been used to fast sinter metals [25,26] and metal oxides [27–30], to the best of the authors knowledge, the highest sintering temperature demonstrated using in-line NIR processing is 450 °C [31]. This temperature value is exceeded in this work, enabling an ultrafast sintering process for NASICON.

A precursor NASICON sol was produced by dissolving ammonium dihydrogen phosphate ($NH_4H_2PO_4$, 0.17g, 1.5 mmol, Fisher Scientific) in $H_2O$ (7.34g) and sodium metasilicate nonahydrate ($Na_2SiO_3.9H_2O$, 0.85g, 3 mmol, Sigma Aldrich) in $H_2O$ (10.00g). The $NH_4H_2PO_4$ solution is then added to the $Na_2SiO_3.9H_2O$ solution and stirred. Tartaric acid ($C_4H_6O_6$, 0.67g, 4.5 mmol, Fisher Scientific) was then added and stirred until dissolved. Zirconyl chloride ($ZrOCl_2.8H_2O$, 0.97g, 3 mmol, Fisher Scientific) was then dissolved in $H_2O$ (30.00g) and stirred > 48 h before filtering through a 0.45μm PTFE syringe filter (Fisher Scientific). The zirconyl chloride solution was added dropwise to the original sample and stirred for 1 h. A 5% solids solution was obtained with a molar ratio of 4:2:2:1 (Na:Zr:Si:P, sodium excess improves $Na_3Zr_2Si_2PO_{12}$ formation[20]).To produce a powder the sol was dried, calcinated and sintered at 1000 °C as detailed by Shimizu [18], producing NASICON powder with negligible secondary phases (Figure 1a). The differential thermal analysis of the dried sol-gel solution (dried at 120 °C, Figure 1b) shows that at the sintering temperature applied, undesired molecules are absent, e.g., physically adsorbed water [32] (dTG peak at 88 °C associated with an endothermic reaction, number 1 in Figure 1b), ammonia from $NH_4H_2PO_4$ [32] and unbounded tartaric acid [33] (dTG peak at 255 °C, associated with an endothermic reaction, number 2 in Figure 1b) and organic molecules coordinated to $Zr^{4+}$ [33] (dTG peaks at 366 °C and 977 °C, both related to exothermic carbon combustion reactions, numbers 3 and 4 in Figure 1b, respectively). The data also shows peaks related to the melting of NaCl around 800 °C and a loss of mass at higher temperatures (number 5 in Figure 1b, dTG at 1055 °C associated with an endothermic peak). It could be related to the evaporation of sodium chloride [17]. The loss of sodium would explain the presence of small secondary phases in the sintered powder (Figure 1a).

To create a thin film, a spray process was developed based on a method used to create compact $TiO_2$ films[34] which is suitable for large area manufacturing. This involved using an artist spray gun (Amazon) to spray the sol, at a nitrogen pressure of 4 bar, onto a fused silica substrate heated to 120 °C (high temperature titanium hot plate), depositing multiple layers to build up the desired thickness. The substrate was pretreated by plasma irradiation ($O_2$ plasma cleaner) for 15 min to improve surface wettability. Once the sample was dried, it was sintered at 1000 °C in air for 3 h. XRD of the resulting film (Figure 1a) shows the film mainly consists of NASICON with some low intensity $SiO_2$ secondary phases (s for Tridymite, orthorhombic silica, COD 9013393, and s' and s'' for cristobalite, tetragonal silica, COD 9001578). Grazing incidence XRD (GI-XRD, Figure 1c) demonstrates that these silica phases were not present on the top surface of the film and are likely the result of the formation of a crystalline silica phase at/near the surface of the substrate. This is confirmed in Figure 1d by the reduction in the intensity of the peaks of these phases with increasing film thickness.

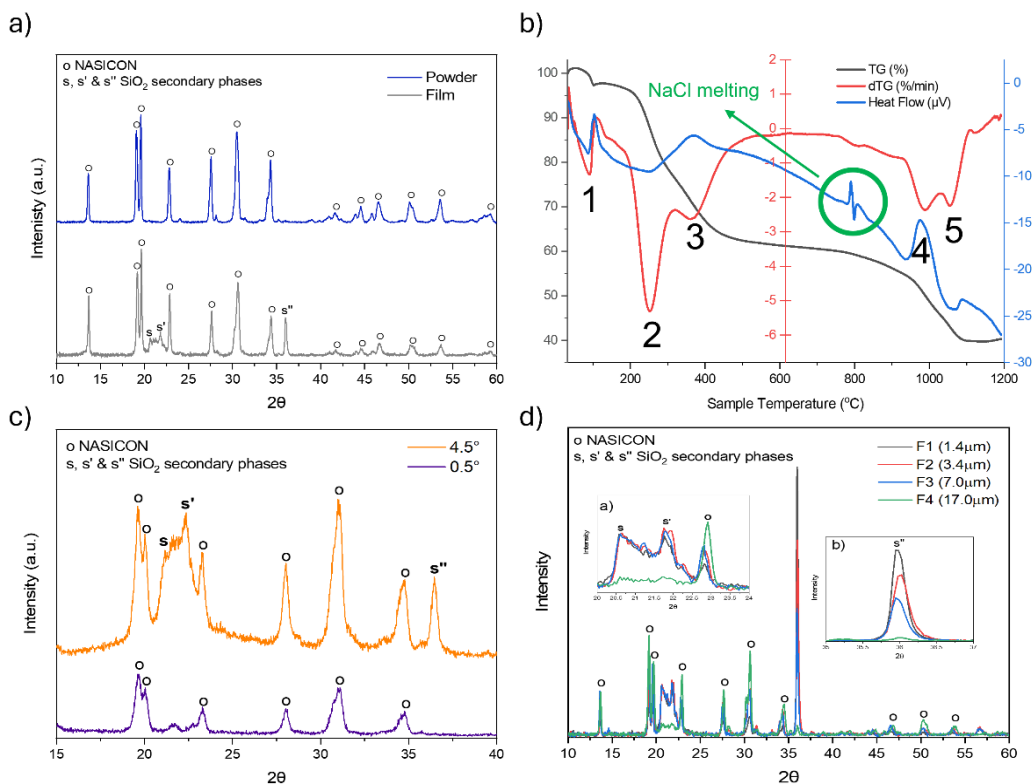

Figure 1: **a)** XRD of NASCION powder and NASICON film on fused silica, **b)** DSC-TGA analysis of dried sol, **c)** GI-XRD of NASICON film at 0.5° and 4.5°, **d)** XRD of NASICON films with different film thicknesses on fused silica substrate.

To reduce the sintering time of the NASICON film from hours to 60 s, near-infrared sintering was employed. This is a process in which the sample is heated radiatively with a peak wavelength intensity of 900-1200nm, (Figure 2a). The equipment used, adphosNIR® 6 × 6 kW ceramic based NIR machine, has heating lamps similar to those employed in previous work (6 x 6 kW output) [31]. Active cooling was applied to the electronic systems along with a quartz bed to allow the sample being processed to reach temperatures of >1000 °C, far exceeding the maximum of 450 °C previously demonstrated with this technology [31]. The lamps also emit longer wavelength infrared radiation (>2000 nm), but its intensity is not as high as in the NIR region.

Both fused silica and the dried sol absorb NIR wavelengths poorly (number 1 and 2, Figure 2b) and therefore direct NIR heating of sprayed sol (NIR oven power at 90% for 60s, 30 layers sprayed) does not result in a temperature high enough to form NASICON (diffractogram A, Figure 2c). It only formed tetragonal (z′ and z′′′, COD 1526427) and monoclinic (z and z′′, COD 2300544) zirconium oxide phases. The crystallisation of these zirconium phases suggests that absorption of longer wavelength IR radiation (>2000nm) happened, e.g., by Si-O and P=O bonds (bonds present in the sample and well-known absorbers of far-infrared radiation). However, the number of photons emitted by the lamps in this region of the spectrum is not enough to reach the fast formation of the NASICON phase.

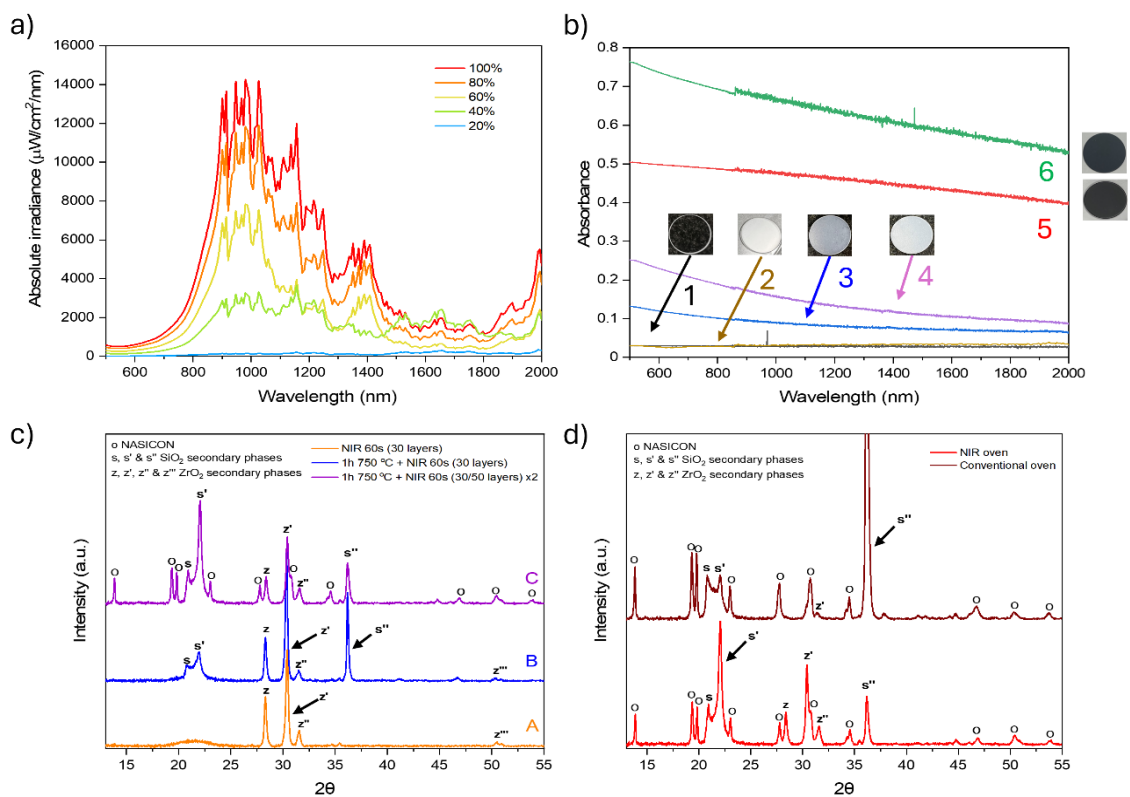

Figure 2: **a)** Output emission of NIR lamps (obtained from[35], more information in supporting content). **b)** NIR absorption of 1: Substrate; 2: 30 sol layers; 3: 30 sol layers, 1h of thermal treatment at 750 °C and NIR sintering at 90% power for 60s; 4: 50 sol layers onto 2, 1h of thermal treatment at 750 °C and NIR sintering at 100% power for 60s; 5: 30 sol layers and 1h of thermal treatment at 750 °C; 6: 50 sol layers onto 2, 1h of thermal treatment at 750 °C. **c)** XRD of NIR sintered samples: A: 30 sol layers directly sintered at 90% power for 60s; B: 30 sol layers, treated at 750 °C for 1h and sintered at 90% power for 60s; C: 50 sol layers onto B, treated at 750 °C for 1h and sintered at 100% power for 60s **d)** XRD of sol sintered in conventional oven and NIR oven.

The tartaric acid contained in the film generates a black carbon-rich intermediate after 1h of heat treatment at 750 °C in air. This material acts as a susceptor that absorbs strongly between 600-2000 nm (number 5, Figure 2b), and therefore heats up the film under NIR radiation. However, after a 750 °C thermal pre-treatment and a NIR sintering step for 60 seconds (heating step at 90% power), the NASICON phase was still not observed (diffractogram B, Figure 2c).

After NIR treatment, the black coloration disappears (number 3, Figure 2b) suggesting that a combustion reaction between carbon and $O_2$ happened. DSC/TGA analysis (Figure 1b) showed that this combustion takes place at temperatures above 950 °C. Furthermore, $SiO_2$ phases appear in the XRD after this treatment with peaks at 20.90°, 22.1° and 36.1° (s, s' and s'' in diffractogram B, Figure 2c). The s' peak (tetragonal silica) is observed starting at 800 °C (thermal stage XRD, Supporting Information, Figure S1a) but the other 2 peaks are not seen until reaching 950 °C (SI, Figure S1b). However, a similar peak intensity for s and s'' is only obtained when sintering at 1000 °C after 15 min (SI, Figures S1c and S1d). This confirms that a temperature of ≥ 1000 °C was reached. Hence, we attribute the non-formation of NASICON to 1) a short sintering time (NASICON phase was observed after 30 min of sintering at 1000 °C in Figure S1d), 2) a too thin film (and, consequently, an insufficient absorption of NIR radiation) and 3) insufficient temperature (i.e., we anticipate that a higher temperature is required to obtain NASICON in such a short time). Interestingly, after depositing a thicker film on top of the aforementioned film (50 more layers, 80 layers in total) and heating it again for 1h at 750 °C, a higher NIR absorbance is obtained (number 6, Figure 2b). Furthermore, after a new NIR sintering process (60s heating step at 100% power), the formation of NASICON occurs (diffractogram C, Figure 2c). We ascribe increasing the thickness to improving the film's ability to absorb NIR radiation and also introducing more carbon (which led to more energy generated in situ during the combustion reaction). This combination of factors enabled the formation of NASICON in a significantly shorter time, i.e., the production time is reduced from 3 h to 60 s. A video of this process is given in Fig S2. The XRD comparison between NIR and conventional oven treatments is shown in Figure 2d showing the same peaks are observed. The difference in the intensity of the NASICON peaks is related to the longer treatment applied for the conventional oven.

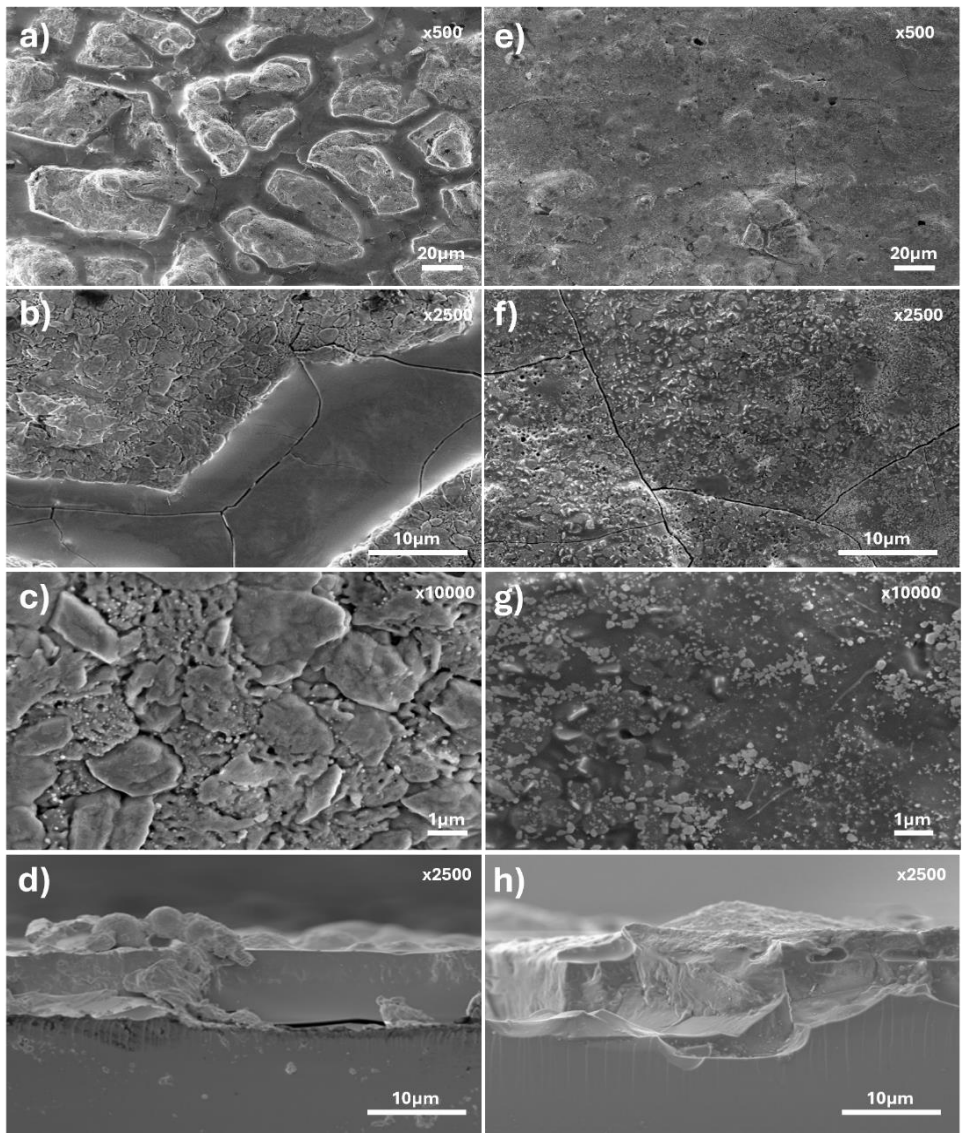

Figure 3.- SEM images from: a-d oven sintered (1000 °C) sprayed NASICON and e-h NIR sintered sprayed NASICON films. Figures 3d and 3h cross sections (oven and NIR, respectively).

In the oven-processed films, the NASICON phase forms into islands connected by $SiO_2$ (identified by EDS, Figure 3a-c), which is attributed to the tetragonal and orthorhombic silica peaks observed in the diffractograms. The NIR sintered NASICON is much more uniform at the micron level (figure 3d) without the two distinct phases. At the submicron scale (figure 3c) secondary phases are evident within the oven sintered NASICON, compared to the NIR sintered microstructure (Figure 3g) where there appear to be two phases but with a much smaller particle size than the oven sintered microstructure. Both films have similar thicknesses of 10 microns (Figure d and h).

In conclusion, we present the first report of NASICON sintered by NIR whilst simultaneously demonstrating a new method to obtain $Na_3Zr_2Si_2PO_{12}$ films by combining the deposition of a water-based sol precursor by spray coating. We combine the power of NIR sintering with the presence of a carbon residue that increases the infrared radiation absorption of the system and allows the crystallisation of the NASICON phase. We believe that this innovative approach can be used to sinter a wide range of ceramic coatings with the potential for faster sintering times, reducing interlayer diffusion and increasing the range of potential substrates.

## CRediT authorship contribution statement

**RMV** experimental, methodology, investigation, formal analysis, write up, conceptualisation, data curation, writing original draft; review and editing. **RG** experimental, data curation, write up, conceptualisation, investigation, review and editing.; **AS** write up; **CD-D** experimental, review and editing. **TD** experimental, formal analysis, review and editing. **SR** experimental, review and editing.; **PJH** supervision, resources, review and editing.; **JB** conceptualisation, methodology, funding, supervision, experimental, resources, writing original draft, review and editing.

## Declaration of Competing Interest

The authors declare that they have no known competing financial interests or personal relationships that could have appeared to influence the work reported in this paper.

## Acknowledgements

We gratefully acknowledge funding from the EPSRC ECR Fellowship NoRESt EP/S03711X/1 (JB, RMV, AS, BG,CD-D), EPSRC/Tata Steel for co-sponsoring an iCASE PhD studentship (Grant number: 2610332) (Voucher # 20000176) for SR and the Sustain Hub (EP/S018107/1) for PJH and HEFCW for the funding capital grant for the STA. The authors would like to thank the access to characterisation equipment to Swansea University Advanced Imaging of Materials (AIM) facility, which was funded in part by the EPSRC (EP/M028267/1) and the European Regional Development Fund through the Welsh Government (80708).

## Supplementary

S1.- Thermal stage XRD results

S2.- Video of NIR sintering process

Characterisation equipment